\documentclass[12pt]{iopart}

\usepackage{iopams}  
\begin{document}

\title[Free energy series for the chromatic polynomial]{Bulk, surface and corner free energy series for the chromatic polynomial on the square and triangular lattices}

\author{Jesper Lykke Jacobsen}

\address{LPTENS, \'Ecole Normale Sup\'erieure, 24 rue Lhomond, 75231 Paris, France}
\address{Universit\'e Pierre et Marie Curie, 4 place Jussieu, 75252 Paris, France}
\ead{jesper.jacobsen@ens.fr}

\begin{abstract}

  We present an efficient algorithm for computing the partition
  function of the $q$-colouring problem (chromatic polynomial) on
  regular two-dimensional lattice strips. Our construction involves
  writing the transfer matrix as a product of sparse matrices, each of
  dimension $\sim 3^m$, where $m$ is the number of lattice spacings
  across the strip. As a specific application, we obtain the large-$q$
  series of the bulk, surface and corner free energies of the
  chromatic polynomial. This extends the existing series for the square
  lattice by 32 terms, to order $q^{-79}$. On the triangular lattice,
  we verify Baxter's analytical expression for the bulk free energy
  (to order $q^{-40}$), and we are able to conjecture exact product
  formulae for the surface and corner free energies.

\end{abstract}

% \maketitle

\section{Introduction}

The $q$-colouring problem consists in assigning to each of the
vertices of a graph $G$ any one of $q$ different colours, in such a
way that adjacent vertices carry different colours. The number of
possible colourings (possibly zero) is known as the chromatic
polynomial $P_G(q)$ of the graph. Although this definition supposes
that $q$ is a positive integer, it is not hard to show
\cite{Birkhoff12,Whitney32} that $P_G(q)$ is indeed a polynomial in
$q$, and so the number of colours can be regarded as a formal
variable.

The history of the $q$-colouring problem is long and interesting, and
we refer the reader to \cite{paper1} for an extensive list of
references. The case where $G$ is a {\em planar} graph has attracted
particular interest. A well-known result is then the four-colouring
theorem \cite{AppelHaken76} which states that $P_G(4) > 0$ for any
planar $G$. Other results exploit the extension of $P_G(q)$ to
non-integer values of $q$. One interesting question is whether there
exists, in the planar case, some $q_{\rm c}$ so that $P_G(q) > 0$ for
all $q \in [q_{\rm c},\infty)$. This statement has been established as
a theorem \cite{BL46} for $q_{\rm c} = 5$, and the so-called 
Birkhoff-Lewis conjecture \cite{BL46} that this extends to
$q_{\rm c} = 4$ is widely believed to be true.

Apart from these general results, the case where $G$ is a {\em regular}
planar lattice---typically square or triangular---has also been
intensively studied. In particular, a long series of papers
\cite{Baxter87,paper1,paper2,paper3,paper4,torus,paper5,paper6} is
devoted to the study of the location and properties of chromatic
zeroes, $P_G(q) = 0$, in the complex $q$-plane for a variety of
boundary conditions. Some of the mechanisms responsible for the
generation of real chromatic zeroes in the region $q \in [0,4]$ have
even been understood analytically \cite{Saleur90,Saleur91}.  One
should also mention that there exists a family of planar graphs with
real chromatic roots converging to $q=4$ from below \cite{Royle05}.

An important technical ingredient in the studies of chromatic zeroes
is the possibility to build up $P_G(q)$, for regular lattice strips of
a given width $m$, by means of a transfer matrix $T$. In order to
elucidate the amazingly intricate behaviour of the chromatic
polynomial in the thermodynamical limit, it is clearly desirable to be
able to access as large $m$ as possible in these computations.

In the present paper we show how to improve the
constructions of \cite{BN82,paper1,paper2,paper3,paper5} by writing $T$ as
a product of {\em sparse} matrices, each of dimension $\sim 3^m$.
This is implemented in an efficient way, that enables us to compute
exactly $P_G(q)$ for rectangles as large as $20 \times 21$. Using the
finite-lattice method \cite{NeefEnting77} this results in series
expansions---in powers of $1/q$---of the bulk, surface and corner free
energies of the chromatic polynomial. To give an idea of the
improvement on existing results, the bulk free energy series on the
square lattice was taken to order $q^{-36}$ in \cite{Bakaev94}, and
very recently extended to order $q^{-47}$ in \cite{paper5}. We here
add a further 32 terms to this latter series, taking it to order
$q^{-79}$. The surface and corner free energies are similarly obtained
to orders $q^{-79}$ and $q^{-78}$.

The case of the triangular lattice---which does not appear to have
been studied previously in terms of series expansions of the free
energy---turns out to be particularly interesting. Baxter
\cite{Baxter86,Baxter87} has shown that for this lattice the chromatic
polynomial possesses an integrable structure. This permitted him to
derive an analytical expression for the bulk free energy, in the form
of an infinite product formula, valid in the region $q \in (-\infty,0]
\cup [4,\infty)$. We obtain in this case the bulk, surface and corner
free energies to order $q^{-40}$. Our results validate Baxter's result
for the bulk free energy, and permit us to conjecture also the exact
infinite product formulae for the surface and corner free energies.

\section{Chromatic polynomial}

We first recall how the chromatic polynomial appears as a special case
of the partition function of the $q$-state Potts model. Given a finite
undirected graph $G=(V,E)$ with vertex set $V$ and edge set $E$, the
Hamiltonian of the Potts model is given by
\begin{equation}
 H = - J \sum_{(ij) \in E} \delta_{\sigma_i,\sigma_j} \,,
\end{equation}
where $J$ is the coupling constant, and a spin $\sigma_i \in
\{1,2,\ldots,q\}$ is defined at each vertex $i \in V$. The Kronecker
delta function $\delta_{\sigma_i,\sigma_j}$ equals 1 if $\sigma_i =
\sigma_j$, and 0 otherwise.

Setting $v = {\rm e}^J-1$, the partition function can be rewritten as
\begin{eqnarray}
 Z_G(q,v) &=& \sum_{\{\sigma\}} \prod_{(ij) \in E}
 \left[ 1 + v \delta_{\sigma_i,\sigma_j} \right] \label{Z1} \\
 &=& \sum_{A \subseteq E} q^{k(A)} v^{|A|} \,, \label{Z2}
\end{eqnarray}
where $k(A)$ is the number of connected components in the subgraph
$G_A = (V,A)$, and $|A|$ is the number of elements in the edge subset
$A \subseteq E$. In the rewriting (\ref{Z2})---known as the
Fortuin-Kasteleyn (FK) representation \cite{FK72}---$q$ does no longer
need to be an integer, and in particular can be used as an expansion
parameter.  For $q \to \infty$, only the term with $A =
\emptyset$---whence $k(A) = |V|$---contributes, and thus one can
develop $Z_G(q,v)$ in powers of $1/q$. The chromatic polynomial
appears as the special case $v=-1$,
\begin{equation}
 P_G(q) = Z_G(q,-1) \,,
\end{equation}
as is evident from (\ref{Z1}).

\section{Transfer matrix formalism}

Thanks to the FK representation (\ref{Z2}), the partition function for
a lattice strip of width $m$ can be built using a transfer matrix $T$
of dimension independent of $q$ \cite{BN82,paper1}. In this section we
briefly review the necessary formalism, with the minor modification
that we rewrite things in terms of $q^{-1}$ rather than $q$, since
this is more convenient for our series expansion purposes. We then
introduce several improvements:
\begin{enumerate}
\item A factorisation of $T$ into sparse matrices that requires only
  the use of non-nearest-neighbour and almost non-nearest-neighbour
  partitions.
 \item An efficient labelling of non-nearest-neighbour partitions by
   natural numbers.
 \item The use of modular arithmetic instead of arbitrary-precision
   integer arithmetic.
\end{enumerate}
The combination of these ingredients will yield an efficient algorithm
that allows, in particular, to add further terms to the series
expansions of \cite{Bakaev94,paper5}.

Let us start by noting that, in view of (\ref{Z2}), $T$ can be taken
to act on states which are partitions of the vertex set
$\{1,2,\ldots,m\}$ corresponding to a single horizontal layer of
spins. The role of $T$ is to propagate the system one lattice spacing
upwards in the vertical direction. The states describe how the
connected components of $G_{\underline{A}} = (V,\underline{A})$
partition the $m$ vertices of a layer, where $\underline{A}$ is the
intersection of $A$ with the set of edges which are {\em below} the
layer under consideration. For instance, when $m=3$, the five possible
states are the partitions
\begin{equation}
 (1)(2)(3) \,, \qquad
 (12)(3) \,, \qquad
 (13)(2) \,, \qquad
 (1)(23) \,, \qquad
 (123) \,.
\end{equation}
We shall see shortly how the information contained in these states
permits us to account for the non-locality of the Boltzmann weight
$q^{k(A)}$.

Our objective is to write the chromatic polynomial on a rectangular
piece, of width $m$ and height $n$, of the square or triangular lattice 
in the form
\begin{equation}
 P_G(q) = \langle e_0 | T_{\rm last} T^{n-1} | e_0 \rangle
 \label{Z_from_T}
\end{equation}
where the basis state $| e_0 \rangle$ corresponds to the all-singleton
partition $(1)(2)\cdots(m)$. Each transfer matrix $T$ adds one layer
to the lattice, consisting of $m-1$ horizontal edges and $m$ vertical
edges---and for the triangular lattice also $m-1$ diagonal edges.
A special matrix $T_{\rm last}$ will be required to deal with the last
layer of $m-1$ horizontal edges. The notation $\langle e_0 |$ means
projection on the all-singleton state, and is not meant to imply the
definition of a scalar product on the space of partitions (although
such a scalar product can indeed be defined).

Let us observe that it would be quite wasteful to construct the matrix
$T$ explicitly. Not only would this require memory resources of the
order of $({\rm dim}\ T)^2$ for its storage, but computing the product
$T | v \rangle$, where $| v \rangle$ represents an arbitrary linear
combination of all possible basis states, would take a time proportional
to $({\rm dim}\ T)^2$. If, on the other hand, we write $T$ as a product
of matrices each corresponding to the addition of a single edge, these
matrices will turn out to be {\em sparse} (with only one or two non-zero
entries per column), and so both time and memory requirements are 
{\em linear} in the dimension of these matrices. In order to proceed,
we first study these single-edge matrices in some detail.

\subsection{Single-edge transfer matrices}

Since we are going to construct a series expansion in powers of $1/q$,
it will be useful to pull out a trivial factor in (\ref{Z2})
and study instead
\begin{equation}
 \tilde{Z}_G(q,v) = q^{-|V|} Z_G(q,v) =
 \sum_{A \subseteq E} q^{k(A)-|V|} v^{|A|} \,.
 \label{Z_modified}
\end{equation}
The lowest-order term, corresponding to $A = \emptyset$, is then $1$.
It is useful to think of the factor $q^{k(A)-|V|}$ in the following way:
whenever the addition of an edge to the set $A$ joins up two connected
component the result is a factor of $q^{-1}$.

The single-edge transfer matrix that adds a horizontal edge between
vertices $i$ and $i+1$ can now be written
\begin{equation}
 {\sf H}_{i,i+1} = 1 + v {\sf J}_{i,i+1} \,,
\end{equation}
where the first (resp.\ second) term represents the possibility of that
edge being absent from (resp.\ present in) the subset $A$. The join
operator ${\sf J}_{i,i+1}$ acts on a partition of $\{1,2,\ldots,m\}$
as the identity if $i$ and $i+1$ belong to the same block. If they
belong to two different blocks, the action of ${\sf J}_{i,i+1}$ is to
join up the blocks and attribute a factor $q^{-1}$.

We now make a remark that will turn out to be very important in what
follows. For the case of the chromatic polynomial (i.e., $v=-1$) ${\sf
  H}_{i,i+1}$ is idempotent. It projects out states in which $i$ and
$i+1$ belong to the same block. It turns out useful to introduce the
normalised projection operator ${\sf P}_{i,i+1}$ which acts on a
partition as the identity if $i$ and $i+1$ belong to different blocks,
and as zero if $i$ and $i+1$ are in the same block. We have then
\begin{equation}
 {\sf H}_{i,i+1} {\sf P}_{i,i+1} = {\sf H}_{i,i+1}
 = \left( {\sf H}_{i,i+1} \right)^2 \,.
 \label{projector}
\end{equation}

The transfer matrix corresponding to a vertical edge between vertices
$i$ and $i'$ can be thought of as the following sequence of operations:
\begin{enumerate}
 \item Insert $i'$ into the partition as a singleton.
 \item Apply ${\sf H}_{i,i'}$.
 \item Eliminate $i$ from the partition.
 \item Relabel $i'$ as $i$.
\end{enumerate}
Alternatively, the vertical edge as can be represented as the operator
\begin{equation}
 {\sf V}_i = v q^{-1} + {\sf D}_i \,,
\end{equation}
where the first (resp.\ second) term represents the situation where
that edge is present in (resp.\ absent from) the subset $A$.
The detach operator ${\sf D}_i$ acts by transforming $i$ into a
singleton. The advantage of the rewriting is that the relabelling
$i' \to i$ is done automatically, but one should keep in mind that
$i$ does {\em not} represent the same vertex on the lattice before
and after the action of ${\sf V}_i$.

Note that although the vertical-edge transfer matrix ${\sf V}_i$ is
related to ${\sf H}_{i,i'}$, it is not a projector in the
chromatic case $v=-1$.

We next show how to combine the observation that ${\sf H}_{i,i+1}$
is a projector with the sparse matrix decomposition of $T$ in order
to achieve a very efficient transfer matrix formalism. Although this
can be done for any regular lattice, we shall discuss only the cases
of the square and triangular lattices.

\subsection{Square lattice}

One can use the operators ${\sf H}_{i,i+1}$ and ${\sf V}_i$ to build
up the lattice, by writing $T$ as a suitable product of such
operators. An obvious construction is to first lay down all the
horizontal edges of a layer, and the add all vertical edges that
connect to the next layer:
\begin{equation}
 T = \left( \prod_{i=1}^m {\sf V}_i \right)
     \left( \prod_{i=1}^{m-1} {\sf H}_{i,i+1} \right)
     \qquad (\mbox{any $v$}) \,.
\end{equation}
Since the next transfer matrix in the product (\ref{Z_from_T}) will
first lay down a row of horizontal edges, we may use (\ref{projector})
and write instead, for $v=-1$,
\begin{equation}
 T = \left( \prod_{i=1}^{m-1} {\sf P}_{i,i+1} \right)
     \left( \prod_{i=1}^m {\sf V}_i \right)
     \left( \prod_{i=1}^{m-1} {\sf H}_{i,i+1} \right)
     \qquad (\mbox{$v=-1$ only}) \,.
 \label{T_as_PVH}
\end{equation}
Although the construction (\ref{T_as_PVH}) takes advantage of the sparse
matrix decomposition, it is not maximally efficient for reasons that
we now point out.

Consider first the case of generic $v$. Successive applications of $T$
to $| e_0 \rangle$, as in (\ref{Z_from_T}), will generate a number of
basis states corresponding to all possible partitions of
$\{1,2,\ldots,m\}$ that are consistent with planarity. These are known
in the literature as non-crossing partitions, and their number is given
by the Catalan number
\begin{equation}
 C_m = \frac{(2m)!}{m!(m+1)!} = 1,2,5,14,42,132,\ldots \,,
\end{equation}
which grows asymptotically as $C_m \sim m^{-3/2} 4^m$ for $m \gg 1$.

For $v=-1$ we can use (\ref{T_as_PVH}), and so once a layer is completed
the number of basis states equals the number of non-crossing
non-nearest-neighbour partitions of $\{1,2,\ldots,m\}$. Their number
is given \cite{paper1} by the Motzkin number $M_{m-1}$, where
\begin{equation}
 M_k = \sum_{j=0}^{\lfloor k/2 \rfloor} {k \choose 2j} C_j
     = 1,1,2,4,9,21,\ldots \,.
\end{equation}
For instance, when $m=4$, the four possible such partitions read
\begin{equation*}
 (1)(2)(3)(4) \,, \qquad
 (1)(24)(3) \,, \qquad
 (13)(2)(4) \,, \qquad
 (14)(2)(3) \,. 
\end{equation*}
The asymptotic behaviour is now $M_k \sim k^{-3/2} 3^k$ for $k \gg 1$.
This looks like a considerable improvement. However, the intermediate
stages that appear when building up the product (\ref{T_as_PVH}) still
require all nearest-neighbour partitions, and so the time necessary
for a complete multiplication with $T$ still grows like $4^m$.

Fortunately, we can write the sparse matrix decomposition of $T$ in a
way which (essentially) only makes use of non-crossing
non-nearest-neighbour partitions. This is accomplished by
utilising the property (\ref{projector}) maximally.
Indeed we can write, for $v=-1$,
\begin{equation}
 T = T_m T_{m-1} \ldots T_3 T_2 T_1
 \label{T_sq_decomp}
\end{equation}
with
\begin{eqnarray}
 T_i &=& \left( \prod_{\stackrel{j=1}{j\neq i}}^{m-1} {\sf P}_{j,j+1} \right)
         {\sf V}_i {\sf H}_{i,i+1} \qquad \mbox{for $i=1,2,\ldots,m-1$}
 \nonumber \\
 T_m &=& \left( \prod_{j=1}^{m-1} {\sf P}_{j,j+1} \right)
         {\sf V}_m \,.
 \label{T_sq_proj}
\end{eqnarray}
After the multiplication by $T_i$, the non-nearest-neighbour
constraint is now in action everywhere, except possibly between sites
$i$ and $i+1$.  The number of partitions where $i$ and $i+1$ do belong
to the same block is $M_{m-2}$, since for such partitions the
contraction of $i$ and $i+1$ into a single point clearly leads to a
non-crossing non-nearest-neighbour partition on $m-1$ points.

The sparse-matrix factorisation (\ref{T_sq_decomp})--(\ref{T_sq_proj})
is a central result of this paper. It implies that, for the square
lattice, the number of basis states needed to compute the chromatic
polynomial using (\ref{Z_from_T}) and (\ref{T_sq_decomp}) is $M_{m-1}$
when a layer has just been completed, and $M_{m-1}+M_{m-2}$ in the
intermediate stages. Both numbers grow like $3^m$.

It remains to describe how the matrix $T_{\rm last}$ lays down the
last layer of horizontal edges. As before this is done one edge at
a time
\begin{equation}
 T_{\rm last} = T_m^{\rm last} T_{m-1}^{\rm last} \ldots
               T_3^{\rm last} T_2^{\rm last} T_1^{\rm last}
 \label{T_last}
\end{equation}
In this product, each factor adds a horizontal bond between $i$ and
$i+1$. Having done that, vertex $i$ and those to the left of it are
of no use any longer, so we might as well detach them as singletons.
This leads to
\begin{equation}
 T_i^{\rm last} = \left( \prod_{j=i+1}^{m-1} {\sf P}_{j,j+1} \right)
         {\sf D}_i {\sf H}_{i,i+1} \qquad \mbox{for $i=1,2,\ldots,m$} \,.
 \label{T_last_i}
\end{equation}
Note that after the action of $T_{\rm last}$, only the all-singleton
state $| e_0 \rangle$ remains. The chromatic polynomial of $G$ appears
as the coefficient of this state, leading finally to (\ref{Z_from_T}).

\subsection{Triangular lattice}

The triangular lattice can be drawn as a square lattice with diagonals.
These diagonals can be constructed, within the transfer matrix approach,
by applying the operators ${\sf H}_{i,i+1}$ just after site $i$ has been
propagated from one layer to the next by means of ${\sf V}_i$, but before
${\sf V}_{i+1}$ has been employed.

We have again the factorisation (\ref{T_sq_decomp}), but now with
\begin{eqnarray}
 T_1 &=& \left( \prod_{j=1}^{m-1} {\sf P}_{j,j+1} \right)
         {\sf V}_1 {\sf H}_{1,2} \nonumber \\
 T_i &=& \left( \prod_{j=1}^{m-1} {\sf P}_{j,j+1} \right)
         {\sf V}_i {\sf H}_{i-1,i} {\sf H}_{i,i+1}
         \qquad \mbox{for $i=2,3,\ldots,m-1$} \nonumber \\
 T_m &=& \left( \prod_{j=1}^{m-1} {\sf P}_{j,j+1} \right)
         {\sf V}_m {\sf H}_{m-1,m} \,.
\end{eqnarray}
Note that, due to the existence of the diagonal edges, the leftmost
factor in each expression is the product over {\em all}
nearest-neighbour projectors. Therefore, the number of basis states
required in any intermediate stage of the construction is $M_{m-1}$.

The expressions (\ref{T_last})--(\ref{T_last_i}) for $T_{\rm last}$
are identical to those for the square lattice.

\subsection{Ordering the states}

At this point, one would be ready to write a transfer matrix for the
chromatic polynomial using only non-crossing non-nearest-neighbour
(NCNNN) partitions ---up to a small subtlety for the square
lattice---as intermediate states, by storing the partitions
encountered using standard hashing techniques.

One can however reduce further---albeit only by a constant
factor---time and memory requirements by establishing a bijection
between the natural numbers $\{1,2,\ldots,M_{m-1}\}$ and the set of
$M_{m-1}$ NCNNN partitions. Using this, the vectors appearing in the
computation of (\ref{Z_from_T}) can be realised simply as standard
arrays, indexed by the natural numbers, and each basis state can be
mapped back and forth between its representation as a natural number
and as a partition on which the operators ${\sf H}_{i,i+1}$, ${\sf
  V}_i$ and ${\sf P}_{i,i+1}$ can be made to act. This idea was
previously implemented in the context of the planar Potts model (i.e.,
for non-crossing partitions) by Bl\"ote and Nightingale \cite{BN82},
and we extend it here by adding the non-nearest-neighbour constraint.

An important step in establishing the bijection is to produce an
ordering of the NCNNN partitions for a given $m$. To this end,
consider the first point in the partition. Two situations may arise:
\begin{enumerate}
\item If point 1 is a singleton, the restriction of the partition to
  the points $\{2,3,\ldots,m\}$ is again NCNNN.
\item If point 1 is not a singleton, let $k \in \{0,1,\ldots,m-3\}$ be
  the {\em smallest} integer so that point $3+k$ belongs to the same block
  as 1. Then the restriction of the partition to points
  $\{2,3,\ldots,2+k\}$ and the restriction to points
  $\{3+k,4+k,\ldots,m\}$ are two independent NCNNN partitions of
  respectively $k+1$ and $m-k-2$ points.
\end{enumerate}
This argument proves in particular the recursion formula (with
$p=m-1$)
\begin{equation}
 M_p = M_{p-1} + \sum_{k=0}^{p-2} M_k M_{p-k-2}
 \label{M_recurs}
\end{equation}
with initial values $M_0 = M_1 = 1$.

Suppose now, as an induction hypotheses, that we have defined an
ordering of NCNNN partitions of fewer than $m$ points. The
ordering of NCNNN partitions of 1 or 2 points is trivial, since in
these cases there is only one such partition.  We can then order
NCNNN partitions of $m$ points as follows:
\begin{enumerate}
\item Place first the partitions in which point 1 is a singleton. The
  ordering of these partitions follows inductively from the ordering
  of the partitions of points $\{2,3,\ldots,m\}$.
\item Place next the partitions in which point 1 is not a singleton,
  in the order of increasing values of the integer $k$ defined above.
  The ordering of the partitions with a fixed value of $k$ follows
  inductively from the ordering of the partitions of points
  $\{2,3,\ldots,2+k\}$ and $\{3+k,4+k,\ldots,m\}$, using the former
  (resp.\ latter) as the most (resp.\ least) significant `bit'.
\end{enumerate}
For instance, the ordering for $m=5$ points reads: \\
\begin{tabular}{lllll}
 (1)(2)(3)(4)(5) \,, &
 (1)(2)(35)(4) \,,   &
 (1)(24)(3)(5) \,,   &
 (1)(25)(3)(4) \,,   & \\
 (13)(2)(4)(5) \,,   &
 (135)(2)(4) \,,     &
 (14)(2)(3)(5) \,,   &
 (15)(2)(3)(4) \,,   &
 (15)(24)(3) \,.
\end{tabular}

\smallskip

The mapping from integers to partitions can now be constructed by
inferring the value of $k$ from (\ref{M_recurs}) and proceeding
recursively. The inverse mapping is constructed similarly by obtaining
$k$ explicitly from the partition.

Finally, for the square lattice, we need a bijection between the
integers $\{1,2,\ldots,M_{m-1}+M_{m-2}\}$ and partitions in which
points $i$ and $i+1$ are allowed to be in the same block. Again this
follows from our ability to order such partitions. To this end, we
place first the $M_{m-1}$ partitions in which $i$ and $i+1$ are not in
the same block, and next the $M_{m-2}$ partitions in which $i$ and
$i+1$ are in the same block. The remaining ingredients of the
bijection can be taken over directly from the NCNNN case.

\subsection{Implementational details}

We wish to compute the chromatic polynomial, normalised as in
(\ref{Z_modified}),
\begin{equation}
 \tilde{P}_G(q) = q^{-|V|} Z_G(q,-1) \,,
\end{equation}
where $G$ is an $m \times n$ rectangular section of the square or
triangular lattice (the latter being considered as a square lattice
with added diagonals). The result is a polynomial of degree $|V|-1$ in
the variable $q^{-1}$ with constant term $1$. To obtain this by the
transfer matrix method, each NCNNN state must have a coefficient
(weight) which is a polynomial of degree $d =
|V|-1$ with integer coefficients. These coefficients can be stored
quite simply as an integer array of length $d + 1$.

Since the goal is to produce a series expansion in powers of $q^{-1}$
of the various free energies, there is actually no need to compute the
whole polynomial $\tilde{P}_G(q)$. We shall see below that one knows
in advance to which degree $d_{\rm max}$ the expansion can be
taken. It suffices therefore to truncate all polynomials intervening
in the coefficients to that order. Such a truncation leads to
non-negligible savings of memory resources.

A practical problem arises because the integer coefficients rapidly
become very large. This is effectively handled by using modular
arithmetic, i.e., by computing the coefficients $\tilde{P}_G(q)$
modulo various primes $p_i$.  Since the transfer process only uses
additions and subtractions, it is convenient to use 32-bit unsigned
integers in the computations and to choose $p_i < 2^{31}$ in order to
avoid overflow before each modulo operation can be carried out. Using
the Chinese remainder theorem, the partial results $\tilde{P}_G(q)
\mbox{ mod } p_i$ can be combined to give $\tilde{P}_G(q) \mbox{ mod }
\prod_i p_i$. Once the latter stabilises upon using further primes we
know that we have arrived at the complete result $\tilde{P}_G(q)$,
with no modulo. The results reported below made necessary the use of
up to 16 different primes.

\section{Results}

We first review briefly how the series expansion for the free energies
can be obtained from the chromatic polynomials on finite $m \times n$
rectangles $G_{m,n}$ by use of the finite-lattice method
\cite{NeefEnting77}. The goal is to obtain a development for the free
energy per vertex in the form
\begin{eqnarray}
  f_{m,n}(q) &\equiv& \frac{\log P_G(q)}{m n} \\
            &=& \left[ \log q + \sum_{k=0}^\infty \frac{a_k}{q^k} \right]
            + \left[ \sum_{k=0}^\infty \frac{b_k}{q^k} \right]
              \left(\frac{1}{m} + \frac{1}{n}\right)
            + \left[ \sum_{k=0}^\infty \frac{c_k}{q^k} \right] \frac{1}{m n} \,,
            \nonumber
\end{eqnarray}
exactly valid, order by order in $q^{-1}$, for all $m$ and $n$
sufficiently large. The quantities within square brackets are referred
to as the bulk, surface and corner free energies, $f_{\rm b}(q)$,
$f_{\rm s}(q)$ and $f_{\rm c}(q)$.

\subsection{Finite-lattice method}

We discuss first the case of the square lattice.  The free energies
introduced above are given by the finite-lattice method
\cite{NeefEnting77} as
\begin{eqnarray}
 f_{\rm b}(q) &=& \log q +
 \sum_{(r,s) \in B(k)} \alpha_k(r,s) \log \tilde{P}_{G(r,s)}(z) + O(z^{2k-3}) \,,
 \label{fb} \\
 f_{\rm s}(q) &=&
 \sum_{(r,s) \in B(k)} \beta_k(r,s) \log \tilde{P}_{G(r,s)}(z) + O(z^{2k-3}) \,,
 \\
 f_{\rm c}(q) &=&
 \sum_{(r,s) \in B(k)} \gamma_k(r,s) \log \tilde{P}_{G(r,s)}(z) + O(z^{2k-3}) \,,
 \label{fc}
\end{eqnarray}
where the set $B(k)$ are $r \times s$ rectangles $G(r,s)$ such that $r
\le s$ and $r + s \le k$. For convenience---and to facilitate the
comparison with previous work \cite{Bakaev94,paper5}---the expansion
variable $q^{-1}$ has here been changed to 
\begin{equation}
 z = \frac{1}{q-1} \,.
 \label{def_z}
\end{equation}
The error term $O(z^{2k-3})$ in (\ref{fb})--(\ref{fc}) is due to
certain classes of convex polygons \cite{NeefEnting77} which do not
fit into any of the rectangles in $B(k)$.

The finite-lattice weights for $f_{\rm b}$ are given by
\cite{NeefEnting77}
\begin{equation}
 \alpha_k(r,s) = (2-\delta_{r,s}) 
 \left( \delta_{r+s,k}-3\delta_{r+s,k-1}+3\delta_{r+s,k-2}-\delta_{r+s,k-3}
 \right)
\end{equation}
while those for $f_{\rm s}$ and $f_{\rm c}$ read \cite{Enting78}
\begin{eqnarray}
 \beta_k(r,s) &=& (2-\delta_{r,s})
 \left(\frac{2-r-s}{2}\delta_{r+s,k}+ 
  \frac{3(r+s)-2}{2}\delta_{r+s,k-1}- \right. \nonumber \\
  & & \left.
  \frac{3(r+s)+2}{2}\delta_{r+s,k-2} +
  \frac{r+s-2}{2}\delta_{r+s,k-3} \right) \,, \\
 \gamma_k(r,s) &=& (2-\delta_{r,s})
  \big( (r-1)(s-1)\delta_{r+s,k}+
  (1+r+s-3rs)\delta_{r+s,k-1}+ \nonumber \\
  & & 
  (3rs+r+s-1)\delta_{r+s,k-2}-
  (r+1)(s+1)\delta_{r+s,k-3} \big) \,.
\end{eqnarray}

\subsection{Square lattice series}

Using the transfer matrix formalism we can compute
$\tilde{P}_{G(r,s)}$ for lattice strips of width $r \le m_{\rm
  max}$. Without resorting to super computer facilities or massive
parallel computations, we have found that $m_{\rm max} = 20$ was
feasible on a standard workstation. This corresponds to a maximum
dimension of the transfer matrices of $M_{19}+M_{18} = 24\,735\,666$
on the square lattice. Setting $k=41$ above, this gives then the free
energies $f_{\rm b}$, $f_{\rm s}$ and $f_{\rm c}$ correctly to order
$z^{78}$. For the case of $f_{\rm b}$ and $f_{\rm s}$ an extra term
can be obtained by adding manually the term $\mu_{2 m_{\rm max}} z^{4
  m_{\max}-1}$ to $f_{\rm b}$ \cite{Bakaev94}, and the term $-m_{\rm
  max} \mu_{2 m_{\rm max}} z^{4 m_{\rm max}-1}$ to $f_{\rm s}$
\cite{paper5}, where \cite{Bakaev94}
\begin{equation}
 \sum_{p=4}^\infty \mu_p x^{2p} =
 x^8 \frac{2-2x^2-x^2\sqrt{1-4x^2}}{(1-4x^2)(2+x^2)} +
 x^{12} \frac{3-4x^2-4\sqrt{1-4x^2}}{(1-4x^2)^2} \,.
\end{equation}
We thus obtain the series to order $z^{79}$ in these cases.

\begin{table}
\begin{center}
\begin{tabular}{|r|r|r|r|} \hline \hline
 $k$ & $\alpha_k$ & $\beta_k$ & $\gamma_k$ \\ \hline \hline
0 & 1 & 1 & 1 \\
1 & 0 & 1 & 0 \\
2 & 0 & 0 & 0 \\
3 & 1 & -1 & 1 \\
4 & 0 & -1 & 0 \\
5 & 0 & 0 & 0 \\
6 & 0 & 1 & 0 \\
7 & 1 & -1 & 4 \\
8 & 3 & -8 & 12 \\
9 & 4 & -16 & 20 \\
10 & 3 & -16 & 28 \\
11 & 3 & -12 & 67 \\
12 & 11 & -41 & 208 \\
13 & 24 & -138 & 484 \\
14 & 8 & -210 & 753 \\
15 & -91 & 47 & 750 \\
16 & -261 & 849 & 679 \\
17 & -290 & 1471 & 2320 \\
18 & 254 & -493 & 10020 \\
19 & 1671 & -8052 & 30548 \\
20 & 3127 & -19901 & 68832 \\
21 & 786 & -19966 & 108744 \\
22 & -13939 & 37556 & 65229 \\
23 & -49052 & 223807 & -236055 \\
24 & -80276 & 508523 & -739289 \\
25 & 21450 & 321314 & 101404 \\
26 & 515846 & -2052462 & 7201383 \\
27 & 1411017 & -8417723 & 26255714 \\
28 & 1160761 & -13374892 & 43505098 \\
29 & -4793764 & 10841423 & -17552274 \\
30 & -20340586 & 112595914 & -291420026 \\
31 & -29699360 & 260687001 & -674637832 \\
32 & 33165914 & 70989018 & 27442 \\
33 & 256169433 & -1341964856 & 4426763291 \\
34 & 495347942 & -4108283969 & 12910062402 \\
\hline \hline
\end{tabular}
\end{center}
\caption{Large-$q$ series for the bulk, surface and corner free energies
of the chromatic polynomial on the square lattice, in terms of the variable
$z=1/(q-1)$. We give the coefficients of
$\exp(f_{\rm b}) = \frac{(q-1)^2}{q} \sum_{k=0}^\infty \alpha_k z^k$, 
$\exp(f_{\rm s}) = \sum_{k=0}^\infty \beta_k z^k$, and
$\exp(f_{\rm c}) = \sum_{k=0}^\infty \gamma_k z^k$.}
\label{tab:sq}
\end{table}

\addtocounter{table}{-1}
\begin{table}
\begin{center}
\begin{tabular}{|r|r|r|r|} \hline \hline
 $k$ & $\alpha_k$ & $\beta_k$ & $\gamma_k$ \\ \hline \hline
35 & -127736296 & -3304416038 & 9737827437 \\
36 & -3068121066 & 14960606999 & -49131891078 \\
37 & -7092358808 & 58237169596 & -184470253912 \\
38 & -1024264966 & 65268280922 & -183956055539 \\
39 & 35697720501 & -162368154719 & 621518352427 \\
40 & 91243390558 & -767619757924 & 2660084937207 \\
41 & 25789733672 & -975329692910 & 3075466954690 \\
42 & -420665229170 & 1872486336165 & -7500763944932 \\
43 & -1089052872105 & 9701425034093 & -34822638005931 \\
44 & -238516756366 & 12262136381593 & -38841312202313 \\
45 & 5101697398582 & -24192583755347 & 104412348649015 \\
46 & 12146149238921 & -118764516172484 & 448320847685638 \\
47 & -598931311074 & -130312353695974 & 421063156900936 \\
48 & -63338329084478 & 346985943489639 & -1507607631713074 \\
49 & -125863684143541 & 1412474002216034 & -5446631048235906 \\
50 & 71523371777335 & 1098760305055003 & -3234717862390686 \\
51 & 790320197578279 & -5183269098276812 & 22683201852542681 \\
52 & 1200257719380100 & -16304598291388925 & 64013086293590143 \\
53 & -1721302814683702 & -5448655170720498 & 8336294391279875 \\
54 & -9694394826555237 & 75813721738614619 & -329171853716322290 \\
55 & -10463394611604378 & 182868102163636571 & -718400221457685146 \\
56 & 29575960135787439 & -33129258404821172 & 310674267788777570 \\
57 & 115472492262862427 & -1051108217408277081 & 4563895897602278816 \\
58 & 85395764100885186 & -2017022924530473193 & 7953356311623456864 \\
59 & -421383549369627730 & 1373959176501915516 & -7763050320815504003 \\
60 & -1343046574902992458 & 13749175266084267367 & -60017851952733958335 \\
61 & -736061882807769534 & 22595026341744522020 & -90527220474610273313 \\
62 & 5291983916641725051 & -23294356475365888405 & 121713573010967365453 \\
63 & 15613586538385554242 & -172293246256852139137 & 768987768617382802911 \\
64 & 8244136015720127003 & -268117182162694379708 & 1123758858144026736112 \\
65 & -61540273845219665190 & 298116598787375322521 & -1553871017012870302628 \\
66 & -186576584391637858059 & 2127014580990734112356 & -9884276597620937431531 \\
67 & -119156912656052312977 & 3423100771562123299025 & -15168542715543012752367 \\
68 & 700439920487427255618 & -3325599173839179924758 & 18496951095018457216592 \\
69 & 2311567965527717105415 & -26543042275747617969190 & 130030721001519590513331 \\
70 & 1784944419294044372980 & -45519479688835787926364 & 210058368547617077899237 \\
\hline \hline
\end{tabular}
\end{center}
\caption{(continued.)}
\label{tab:sq1}
\end{table}

\addtocounter{table}{-1}
\begin{table}
\begin{center}
\begin{tabular}{|r|r|r|r|} \hline \hline
 $k$ & $\alpha_k$ & $\beta_k$ & $\gamma_k$ \\ \hline \hline
71 & \small -8190295787656754694494 & \small 36371250229750107503644 & \small -228752490963172501272677 \\
72 & \small -29120128759101310463081 & \small 336827820960866947369728 & \small -1729404450220843332781816 \\
73 & \small -24509224734165974718725 & \small 600304759572952426851130 & \small -2809877317417695436535077 \\
74 & \small 99609960301714320809469 & \small -429784892999554830063487 & \small 3052097912541295946004847 \\
75 & \small 363523561198034374612305 & \small -4289105251916055148255439 & \small 22691488586148969734635242 \\
76 & \small 307076608628563447831354 & \small -7658381117547376351931511 & \small 35816140480454963887649892 \\
77 & \small -1234243320099804373986722 & \small 5508060829297214833104965 & \small -41473851023699412644120745 \\
78 & \small -4471768398561919868797416 & \small 54201975484843398706779272 & \small -291274582715872945456529116 \\
79 & \small -3688741741221487636404873 & \small 95450369797304235548078036 & \small \\
\hline \hline
\end{tabular}
\end{center}
\caption{(continued.)}
\label{tab:sq2}
\end{table}

The results for all three series are given in Table~\ref{tab:sq}.
They extend those of \cite{paper5} by 32 new terms.

It is possible to analyse%
\footnote{We are grateful to J.~Salas for providing the analysis
  mentioned in this paragraph.}
the extended series using the method of differential approximants.
The goal is to obtain the locations $z_\ell$ and exponents
$\lambda_\ell$ of the singularities of the free-energy series, which
is supposed to behave like $\sim (z-z_\ell)^{\lambda_\ell}$ as $z \to
z_\ell$.  Proceeding exactly as described in section 5.5.3 of
\cite{paper5}, one obtains for the bulk free energy the
first-order ($K=1$) approximants
\begin{equation}
\begin{tabular}{ll}
 $z_1 = 0.24(4) \pm 0.40(5) i$ & $\lambda_1 = -1.8(7)$ \\
 $z_2 = 0.64(7)$               & $\lambda_2 = -1.6(7)$ \\
 $z_3 = -1.00(1)$              & $\lambda_3 = 0.99(6)$ \\
\end{tabular}
\end{equation}
while the second-order ($K=2$) approximants read
\begin{equation}
\begin{tabular}{ll}
 $z_1 = 0.24(4) \pm 0.40(3) i$ & $\lambda_1 = -2.4(7)$ \\
 $z_2 = 0.62(10)$              & $\lambda_2 = -1.3(16)$ \\
\end{tabular}
\end{equation}
Somewhat disappointingly, these results hardly improve on those
reported in \cite{paper5}, despite of the substantial extensions of
the series. The same is true for the reanalysis of the surface and
corner free energies.

\subsection{Triangular lattice series}

For the triangular lattice, we can use exactly the same finite-lattice
weights as for the square lattice. Taking again $m_{\rm max} = 20$ and
$k=41$ corresponds to a maximum dimension of the transfer matrices of
$M_{19} = 18\,199\,284$. The error term in (\ref{fb})--(\ref{fc}) is
now $O(z^k)$, and we know of no special tricks for obtaining an extra
term. Therefore, we obtain in this case the three series to order
$z^{40}$ included.

In a couple of important papers, Baxter \cite{Baxter86,Baxter87} has
shown that the triangular-lattice chromatic polynomial is an
integrable system. In particular, he has set up the Bethe Ansatz
equations and derived three different analytical expressions for the
bulk free energy $f_{\rm b}$. The question which one of the three
expressions dominates for a given real---or even complex---value of
$q$ turns out to be very intricate \cite{Baxter87}, and has recently
been the subject of a careful reanalysis \cite{paper3}. The expression
which is dominant for real $q \in (-\infty,0] \cup [4,\infty)$ turns
out to be
\begin{equation}
 {\rm e}^{f_{\rm b}} = -\frac{1}{x} \prod_{j=1}^\infty
 \frac{(1-x^{6j-3})(1-x^{6j-2})^2(1-x^{6j-1})}
      {(1-x^{6j-5})(1-x^{6j-4})(1-x^{6j})(1-x^{6j+1})}
 \label{efb}
\end{equation}
where the previous expansion variable (\ref{def_z}) has now been
traded for $x$ which is defined by
\begin{equation}
 q = 2 - x - \frac{1}{x} \,.
\end{equation}
As expected, our series expansion validates (\ref{efb}) to order $x^{40}$
included.

The analytic computation of $f_{\rm s}$ and $f_{\rm c}$ was not
attempted in \cite{Baxter86,Baxter87}. Presumably $f_{\rm s}$ could be
derived by a substantial adaptation of Baxter's work in order to
impose non-periodic boundary conditions in his transfer matrix
formalism.  On the other hand, the derivation of $f_{\rm c}$ appears
to be a much harder problem, since this would require precise
knowledge of the correspondence between boundary states in the Bethe
setup and the initial lattice model, as well as the ability to compute
scalar products in the Bethe formalism.

The series expansions that we obtain for $f_{\rm s}$ and $f_{\rm c}$
read as follows
\begin{eqnarray}
 {\rm e}^{f_{\rm s}} &=& 1 - 2 x + x^2 + 2 x^3 - 2 x^4 - x^6 + 2 x^7 + 2 x^8 -
 4 x^9 - 2 x^{11} + 7 x^{12} - \nonumber \\
 & & 8 x^{14} + 4 x^{15} - 5 x^{16} + 12 x^{17} - 12 x^{19} + 4 x^{20} -
 10 x^{21} + 24 x^{22} - \nonumber \\
 & & 2 x^{23} - 19 x^{24} + 2 x^{25} - 15 x^{26} + 46 x^{27} - 10 x^{28} - 
 24 x^{29} - x^{30} - \nonumber \\
 & & 24 x^{31} + 77 x^{32} - 22 x^{33} - 25 x^{34} - 16 x^{35} - 35 x^{36} + 
 118 x^{37} - \nonumber \\
 & & 41 x^{38} - 14 x^{39} - 45 x^{40} + O(x^{41}) \,, \label{efs_ser} \\
 {\rm e}^{f_{\rm c}} &=& 1 + x - 3 x^3 - 5 x^4 - 6 x^5 + 2 x^6 + 11 x^7 +
 21 x^8 + 9 x^9 - 12 x^{10} - \nonumber \\ 
 & & 41 x^{11} - 33 x^{12} + 72 x^{14} + 82 x^{15} + 40 x^{16} -
 102 x^{17} - 169 x^{18} - \nonumber \\
 & & 139 x^{19} + 103 x^{20} + 273 x^{21} + 319 x^{22} - 50 x^{23} -
 398 x^{24} - 612 x^{25} - \nonumber \\
 & & 111 x^{26} + 517 x^{27} + 1066 x^{28} + 454 x^{29} - 537 x^{30} -
 1686 x^{31} - \nonumber \\
 & & 1067 x^{32} + 398 x^{33} + 2500 x^{34} + 2079 x^{35} + 76 x^{36} -
 3521 x^{37} - \nonumber \\ 
 & & 3641 x^{38} - 1138 x^{39} + 4643 x^{40} + O(x^{41}) \,. \label{efc_ser}
\end{eqnarray}
It is seen that the coefficients are numerically much smaller than
those for the square lattice (see Table~\ref{tab:sq}). This in itself
is not surprising, since the radius of convergence is $1$ in the $x$
variable (triangular-lattice case), while that of the $z$ variable
(square-lattice case) is $\simeq \frac12$ (and presumably exactly
$\frac12$ in view of \cite{Baxter82}). What is truly surprising is
that the coefficients are integers! It thus looks reasonable to try to
conjecture exact expressions.

In analogy with other problems where both bulk and surface
properties have been worked out, one would expect the periodicity of 6
in the product formula (\ref{efb}) to be doubled to 12 in the
corresponding surface quantities. Taking this as a clue suggests trying
for ${\rm e}^{f_{\rm s}}$ and ${\rm e}^{f_{\rm c}}$ product formulae of the
form
\begin{equation}
 \prod_{j=1}^\infty \prod_{k=0}^{11} (1-x^{12j-k})^{\alpha_k}
\end{equation}
for some exponents $\alpha_k$ to be determined. Knowledge of the
coefficients of $x^1,\ldots,x^{12}$ from
(\ref{efs_ser})--(\ref{efc_ser}) allows us to determine
$\alpha_0,\ldots,\alpha_{11}$; we can then test our conjectural formula
against the coefficients of $x^{13},\ldots,x^{40}$.

In this way we have obtained the conjecture
\begin{equation}
 {\rm e}^{f_{\rm s}} = \prod_{j=1}^\infty \left(
 \frac{(1-x^{12j-11})(1-x^{12j-6})(1-x^{12j-5})(1-x^{12j-4})}
      {(1-x^{12j-9})(1-x^{12j-8})(1-x^{12j-7})(1-x^{12j-2})} \right)^2
 \label{efs}
\end{equation}
for the surface free energy, and
\begin{eqnarray}
 {\rm e}^{f_{\rm c}} &=& \prod_{j=1}^\infty
 (1-x^{12j-11})^{-1} (1-x^{12j-10}) (1-x^{12j-9})^3 (1-x^{12j-8})^2 \times
 \nonumber \\
 & & \quad \ (1-x^{12j-7})^4 (1-x^{12j-6})^{-3} (1-x^{12j-5}) (1-x^{12j-4})^{-3} \times
 \nonumber \\
 & & \quad \ (1-x^{12j-3})^5 (1-x^{12j-2})^2 (1-x^{12j-1})^2 (1-x^{12j})
 \label{efc}
\end{eqnarray}
for the corner free energy. The validation of (\ref{efs})--(\ref{efc})
by $40-12=28$ independent checks means that both conjectures are
established beyond any reasonable doubt.

\subsection{Limits $q \to 0$ and $q \to 4$}

It is obvious that the product formulae (\ref{efb}), (\ref{efs}) and
(\ref{efc}) converge for $|x| < 1$, since this is true for each
individual factor in the product (by taking logarithms). It remains to
discuss the limiting cases $x \to 1$ (i.e., $q \to 0$) and $x \to -1$
(i.e., $q \to 4$). The latter case---four-colouring the vertices of
the triangular lattice---can be shown to be equivalent to other
physically interesting problems, such as three-colouring the edges of
the hexagonal lattice \cite{Baxter70,MooreNewman00}, or fully-packed
loops on the hexagonal lattice with fugacity $n_{\rm loop}=2$
\cite{KondevHenley94}.

In the $x \to 1$ case, we write $x = 1 - \epsilon$ for $\epsilon \ll 1$.
Then, to leading order, we have for each factor in the products
$1-x^\alpha \simeq \alpha \epsilon$. This leads to
\begin{eqnarray}
 \lim_{x \to 1} {\rm e}^{f_{\rm b}} &=&
 - \prod_{j=1}^\infty \frac{(6j-3)(6j-2)^2(6j-1)}{(6j-5)(6j-4)(6j)(6j+1)}
 \nonumber \\
 &=& \frac{3^{3/2} \Gamma \left( \frac13 \right)^9}{(2\pi)^5} 
 \simeq 3.770919\cdots \,, \\
 \lim_{x \to 1} {\rm e}^{f_{\rm s}} &=&
 \prod_{j=1}^\infty \left[ \frac{(12j-11)(12j-6)(12j-5)(12j-4)}
 {(12j-9)(12j-8)(12j-7)(12j-2)} \right]^2 \nonumber \\
 &=& \frac{1}{2\pi} \left[
 \frac{\Gamma \left( \frac{5}{12} \right)
       \Gamma \left( \frac{1}{4} \right)}
      {\Gamma \left( \frac{1}{6} \right)} \right]^2
 \simeq 0.305639\cdots \,, \\
 \lim_{x \to 1} {\rm e}^{f_{\rm c}} &=& 0 \,.
\end{eqnarray}

In the $x \to -1$ case, we write $x = -1 + \epsilon$ for $\epsilon \ll
1$.  When $\alpha$ is even, we have then $1-x^\alpha \simeq \alpha
\epsilon$ as before. And when $\alpha$ is odd, $1-x^\alpha \simeq 2$ to
leading order. This leads to
\begin{eqnarray}
 \lim_{x \to -1} {\rm e}^{f_{\rm b}} &=&
 \prod_{j=1}^\infty \frac{(3j-1)^2}{(3j-2)(3j)} 
 = \frac{3 \Gamma \left( \frac13 \right)^3}{4 \pi^2}
 \simeq 1.460998\cdots \,, \\
 \lim_{x \to -1} {\rm e}^{f_{\rm s}} &=&
 \prod_{j=1}^\infty \left[ \frac{(6j-3)(6j-2)}{(6j-4)(6j-1)} \right]^2
 = 4^{1/3} \simeq 1.587401\cdots \,, \\
 \lim_{x \to -1} {\rm e}^{f_{\rm c}} &=& \infty \,.
\end{eqnarray}
It is remarkable that $f_{\rm c}$ diverges in both limits $x \to \pm 1$.

\section{Discussion}

In this paper we have used the finite-lattice method to obtain
large-$q$ series for the bulk, surface and corner free
energies---denoted $f_{\rm b}$, $f_{\rm s}$ and $f_{\rm c}$---for
chromatic polynomials on regular two-dimensional lattices. Long series
could be obtained due to a technical improvement of the transfer
matrix method, which is specific to the antiferromagnetic
zero-temperature limit of the Potts model, and amounts to obtaining a
factorisation of $T$ in terms of sparse matrices, each of dimension
$\sim 3^m$, where $m$ is the strip width. Although this applies in
principle to any regular lattice, we have focussed here on the most
commonly studied lattices (square and triangular).

For the square lattice the improvement allowed us to add 32 new terms
to the existing series \cite{paper5}. Unfortunately, the reanalysis of
the singularities of this series, using the method of differential
approximants, did not lead to any significant improvement over
\cite{paper5}.  The focus in this paper was therefore rather on the
triangular lattice, where the series permitted us to conjecture the
exact analytical formulae (\ref{efs})--(\ref{efc}) for $f_{\rm s}$ and
$f_{\rm c}$, and to confirm the correctness of Baxter's
\cite{Baxter86,Baxter87} result (\ref{efb}) for $f_{\rm b}$.

It should be noticed that going to $m_{\rm max}=20$ leads to $79$
terms in the square-lattice series, but only to $40$ terms for the
triangular lattice. Longer series could be obtained in the latter case
by using a version of the finite-lattice method which is specifically
adapted to the triangular lattice \cite{EntingTri}. However, since our
series have enabled us to conjecture the exact (infinite-order)
results (\ref{efs})--(\ref{efc}) the motivation for extracting further
terms of the series is admittedly rather scarce.  Indeed, the series
reported here already provide $40-12=28$ independent checks of the
proposed expressions (\ref{efs})--(\ref{efc}).

It would be interesting to elucidate the physical mechanism by which
$f_{\rm c}$ diverges in the limits $q \to 0$ and $q \to 4$.

It is regrettable that the series obtained here do not provide any
information about the regions where the chromatic polynomial is a
critical system---that is, $q \in [0,3]$ for the square lattice
\cite{Baxter82,Saleur90,Saleur91,paper1,paper2,torus} and $q \in
[0,4]$ for the triangular lattice
\cite{Baxter86,Baxter87,paper3,torus}. The reason is that $q=\infty$
seems to be the only viable expansion point for the finite-lattice
method.

It is possible that the fact that non-nearest neighbour constraint can
be made local conceals some deeper representation theoretical meaning.
It is well-known that the generic representation theory of the
Temperley-Lieb algebra, which is central to the algebraic formulation
of the planar Potts model, undergoes profound modifications when $q =
4 \cos(\pi/(p+1))^2$ with $p$ integer (or rational)
\cite{PasquierSaleur}. It might be that similar modifications occur
upon taking the particular value $v=-1$ for the temperature variable.
It should be noticed that some steps toward an algebraic formulation of
the chromatic polynomial have been taken in \cite{Fendley}.

\section*{Acknowledgements}

This work was supported by the Agence Nationale de la Recherche (grant
ANR-06-BLAN-0124-03). The author expresses his gratitude to J.\ Salas
and A.D.\ Sokal for a careful reading of the manuscript; and in
particular to J.\ Salas for reanalysing the square-lattice series
along the lines of \cite{paper5}.

\end{document}